\documentclass[%
 reprint,
superscriptaddress,
 amsmath,amssymb,
 aps,
prl,
]{revtex4-2}

\usepackage{graphicx}
\usepackage{dcolumn}
\usepackage{bm}
\usepackage{xcolor}

\begin{document}

\title{Particle scale anisotropy controls bulk properties in sheared granular materials}
\author{Carmen L. Lee}
\affiliation{Department of Physics, North Carolina State University, Raleigh, North Carolina, 27695, USA}

\author{Ephraim Bililign}%
\affiliation{Department of Physics, North Carolina State University, Raleigh, North Carolina, 27695, USA}
\affiliation{James Franck Institute and Department of Physics, University of Chicago, Chicago, IL, USA}

\author{Emilien Az\'ema}
\affiliation{LMGC, Universit\'e de Montpellier, CNRS, Montpellier, France}
\affiliation{Department of Civil, Geological, and Mining Engineering, Polytechnique Montréal, Montréal, Canada.}
\affiliation{Institut Universitaire de France (IUF), Paris, France}

\author{Karen E. Daniels}
\affiliation{Department of Physics, North Carolina State University, Raleigh, North Carolina, 27695, USA}

\date{\today}

\begin{abstract}
The bulk dynamics of dense granular materials arise  through a combination of particle-scale and mesoscale effects.
Theoretical and numerical studies have shown that collective effects are created by particle-scale anisotropic structures such as grain connectivity (fabric), force transmission, and frictional mobilization, all of which
influence bulk properties like bulk friction and the stress tensor through the Stress-Force-Fabric (SFF) relationship. To date, establishing the relevance of these effects to laboratory systems has remained elusive due to the challenge of measuring both normal and frictional contact forces at the particle scale. In this study, we perform experiments on a sheared photoelastic granular system in an quasi-2D annular cell. During these experiments, we measure particle locations, contacts, and normal and frictional forces vectors during loading. We reconstruct the angular distributions of the contact and force vectors, and extract the corresponding emergent anisotropies for each of these metrics. Finally, we show that the SFF relation quantitatively predicts the relationship between particle scale anisotropies, the stress tensor components, and the bulk friction coefficient, capturing even transient behaviors.
As such, this method shows promise for application to other dense particulate systems where fabric anisotropy can provide a useful measure of bulk friction.
\end{abstract}

\maketitle

Foams, emulsions, colloidal suspensions, granular media, cellular tissues, and other amorphous materials all exhibit bulk phenomena that arise due to collective interactions, complicating the development of continuum models. Granular materials in particular have a poor separation of length scales, meaning that particle scale and mesoscale interactions heavily influence bulk properties~\cite{tordesillas_bridging_2004, keys_measurement_2007, shahin_quantifying_2022, goldhirsch_scales_1999, liu_spongelike_2021}. Although bulk measures like stress, strain, or the bulk friction coefficient are well defined, the measured values vary because these properties are determined by the loading history. At best, continuum measures serve as ensemble information on the state of the granular material as a whole at that snapshot in time, rather than being a material property. In particular, it is well established that material strength arises from the buildup of various anisotropic structures at the particle scale, induced by grain connectivity (fabric), force transmission, and frictional mobilization~\cite{radjai_modeling_2017, borzsonyi_granular_2013,li_fabric_2014}.

\begin{figure}[b]
\includegraphics[width=\columnwidth]{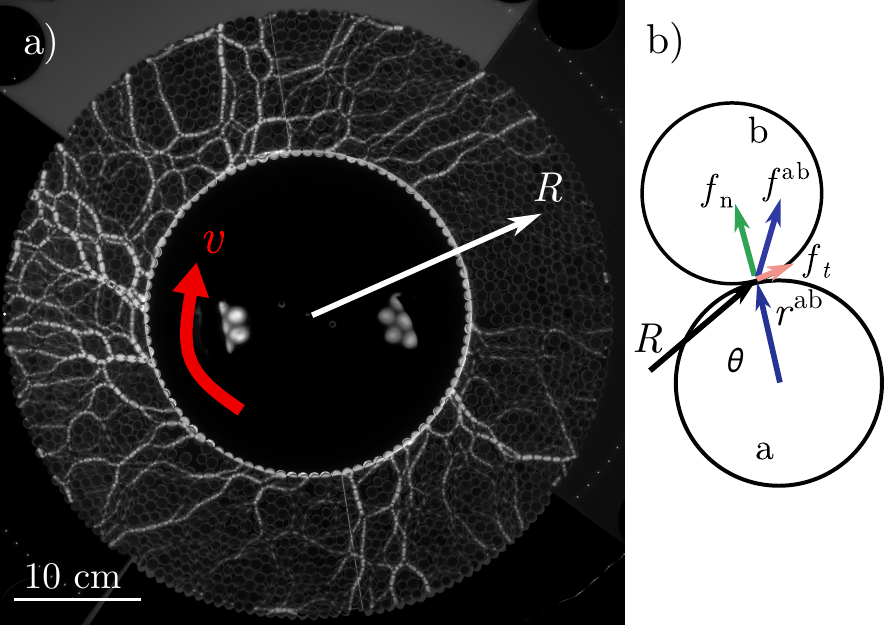}
\caption{\label{fig:history} (a) Top view of photoelastic particles viewed through a darkfield polariscope (white = higher stress) inside of an annular shear cell with inner boundary rotating at a constant velocity $v$ and fixed outer boundary. Each particle provides a measurement of its own contact forces, which we measure relative to the radial polar coordinate $R$.
(b) Sketch of the interparticle coordinate system, with the contact position vector $r^{ab}$ pointing from the center of particle $a$ to the contact point with particle $b$. The force exerted by particle $a$ on $b$ is given by $f^{ab}$; this can decomposed into normal $f_n$ and tangential $f_t$. The angle $\theta$ defines the relative angle between $r^{ab}$ and $R$ for the branch vector and position vector of the contact between particles $a$ and $b$.}
\end{figure}

Because fabric strongly influences bulk measures, even when the particles undergo identical loading history, jammed systems additionally exhibit anisotropy in the shear modulus~\cite{dagois-bohy_soft-sphere_2012, tighe_force_2010}. 
In rheological contexts, bulk friction coefficients have been shown to depend on the combination of fabric and force~\cite{da_cruz_rheophysics_2005,azemaInternalStructureInertial2014}. Theoretical micromechanical models have been developed to further understand granular material from their rheology~\cite{jop_constitutive_2006, kamrin_nonlocal_2012, berzi_granular_2024} to stability and jamming~\cite{cates_jamming_1998, chang_micromechanical_2011} and many studies have quantified the anisotropic structures appearing while under load ~\cite{majmudar_contact_2005, wang_strength_2017, shi_fabric_2018, hurley_quantifying_2016, bi_jamming_2011}.

All of these challenges can be addressed with a framework first pioneered by \citet{rothenburg_analytical_1989}. The Stress-Force-Fabric (SFF) relation connects the bulk stress tensor to particle scale anisotropies in contact orientation, contact vector length, and force vectors;  it has been verified extensively in DEM simulations of granular materials of varying complexity to understand the role of particle scale structure in bulk measurements~\cite{azema_force_2007,azema_shear_2017,binaree_combined_2020,zhou_microscopic_2024,liu_how_2022}. 
Further work by~\citet{li_stressforcefabric_2013} rigorously derived the SFF relation using a directional statistics approach and established its key assumptions. However, the SFF relation has seen little attention in laboratory systems because it requires measurements of individual particle contacts and vector forces. Recent developments in photoelastic techniques~\cite{abed_zadeh_enlightening_2019, daniels_photoelastic_2017, kollmer_photo-elastic_nodate}, combined with high-resolution photography, now allow for force and contact resolution suitable to test the SFF relation using photoelastic granular assemblies.

 In this study, we use photoelastic disks ~\cite{abed_zadeh_enlightening_2019,daniels_photoelastic_2017} arranged in a single layer within an annular shear cell (Fig.~\ref{fig:history}(a)) to map emergent particle scale anisotropic structures to bulk friction and the stress tensor using the SFF relation and further apply it to capturing dynamic behaviour. In doing so, we address the open question of the influence of particle scale anisotropy during granular dynamics and show the first laboratory demonstration of the SFF relation. The particles begin in an unstrained state with no observable forces and the inner boundary is rotated to shear the assembly.  During these experiments, we record particle locations, contacts, and normal and frictional forces vectors for each particle and calculate the stress tensor and anisotropies. In the companion article we use the SFF relation to study the role of loading condition on the evolution of bulk properties~\cite{lee_preprint2}

We performed experiments in a constant-area quasi-two-dimensional annular shear cell containing $N_p$ = 1810 bidisperse (Vishay PhotoStress PSM-4) circular disks with an equal fraction of small particles with radii $R_s$ = 4.5 mm and large particles with radii $R_l$ = 5.5 mm. The particles are dusted with baking powder to reduce basal friction between the particles and the supporting sheet, and the interparticle friction coefficient is estimated to be 0.3. The inner and outer boundaries of the annulus have particle-sized cavities to promote gripping between the boundaries and the particles. 
The inner boundary of the annulus is rotated at a constant rate of 1 rotation every 2.5 hours and the packing undergoes simple shear.  Further details of the annulus can be found in previous works~\cite{brzinski_sounds_2018, fazelpour_controlling_2023}. Using the annular strain rate $\Dot{\gamma} = \frac{1}{2} \left(\frac{\partial v}{\partial r} - \frac{v}{r}\right)$, and integrating over time with appropriate boundary conditions, we calculate the strain $\epsilon$. The packing is illuminated with both red unpolarized light (particle-detection) and green circularly polarized light (darkfield polariscope). 
Images are taken  at a frame rate of 5 seconds per frame with a Nikon D850 camera.  We extract particle position, contact and vector force data from the images of the particles as outlined in previous works ~\cite{abed_zadeh_enlightening_2019, daniels_photoelastic_2017} by using the open source code, PhotoElastic Grain Solver~\cite{kollmer_photo-elastic_nodate} that we modified to detect particle-particle contacts with force magnitudes with a resolution of $5 \times 10^{-5}$ N. Further details of the analysis can be found in the companion article~\cite{lee_preprint2}.

From the particle scale information, we calculate the stress tensor in two dimensions using the micro-structural definition~\cite{kruyt_micromechanical_2014, bagi_microstructural_1999}. We compute the two dimensional stress tensor via summation of the outer product of the force $\vec{f}$ and contact position vector $\vec{r}$~\cite{bagi_microstructural_1999, li_stressforcefabric_2013, rothenburg_analytical_1989}:

\begin{equation}
    \sigma_{ij} = \frac{1}{S} \sum_{a \neq b} \vec{r^{ab}_i} \vec{f^{ab}_j},
    \label{eq:stresstensor}
\end{equation}
where in our two dimensional system, $S$ is the area of the annulus. 
Here we follow classical notation,
where the contact position vector points from the center of the grain to the contact point, and the force vector points from the contact point in the direction of the contact force. Figure~\ref{fig:history}(b) shows a schematic of the vectors for two sample particles. From this formulation of the stress tensor, we are able to calculate the bulk friction coefficient,
\begin{equation}
    \mu = \frac{\tau}{P} = \frac{\sqrt{(\sigma_{11}-\sigma_{22})^2+4\sigma_{12}\sigma_{21}}}{(\sigma_{11} + \sigma_{22})/2},\label{eq:mu}
\end{equation}
where the numerator is the shear component of the stress tensor, and the denominator is the mean pressure.

To consider the anisotropies in a granular material, the SFF relation relies on decomposing the stress tensor into angular distributions according to the 
unit contact vector $\hat n$. In 2D, the unit contact vector is described by a single angle $\theta$, and 
assuming the thermodynamic limit, the stress tensor can be written as:

\begin{equation}
    \sigma_{ij} = \frac{N_c}{S} \oint_\theta E^c({\theta}) \left< \ell_i f_j \right> (\theta) d\theta,\label{eq:stressexpand}
\end{equation}
 where $N_c$ is the total number of contacts in $S$, $\vec \ell$ is the  branch vector (the vector joining the centers of two contacting particles), and $E^c({\theta}$) is the probability distribution function (pdf) of a contact pointing along the direction $\theta$.

Because our system is composed of round, nearly monodisperse particles, the branch vector can be approximated as $\vec \ell = d \hat n$, with $d$ the mean grain diameter, and therefore there is no  
preferred orientation to the branch vector~\cite{azema_shear_2017}.  
Because, in general, the vector forces and branches are statistically independent, we approximate $\langle \ell_i f_j \rangle (\theta) \sim d n_i (\theta) \langle f_j \rangle(\theta)$. The angular distribution of the mean vector force $\langle f_j \rangle(\theta)$ can be decomposed according the mean angular distribution of normal $\langle f_n \rangle(\theta)$ and tangential $\langle f_t \rangle(\theta)$ forces as $\langle f_j \rangle(\theta) = \langle f_n \rangle(\theta)n_j + \langle f_n \rangle(\theta)t_j$.

\begin{figure*}
\includegraphics[width=\textwidth]{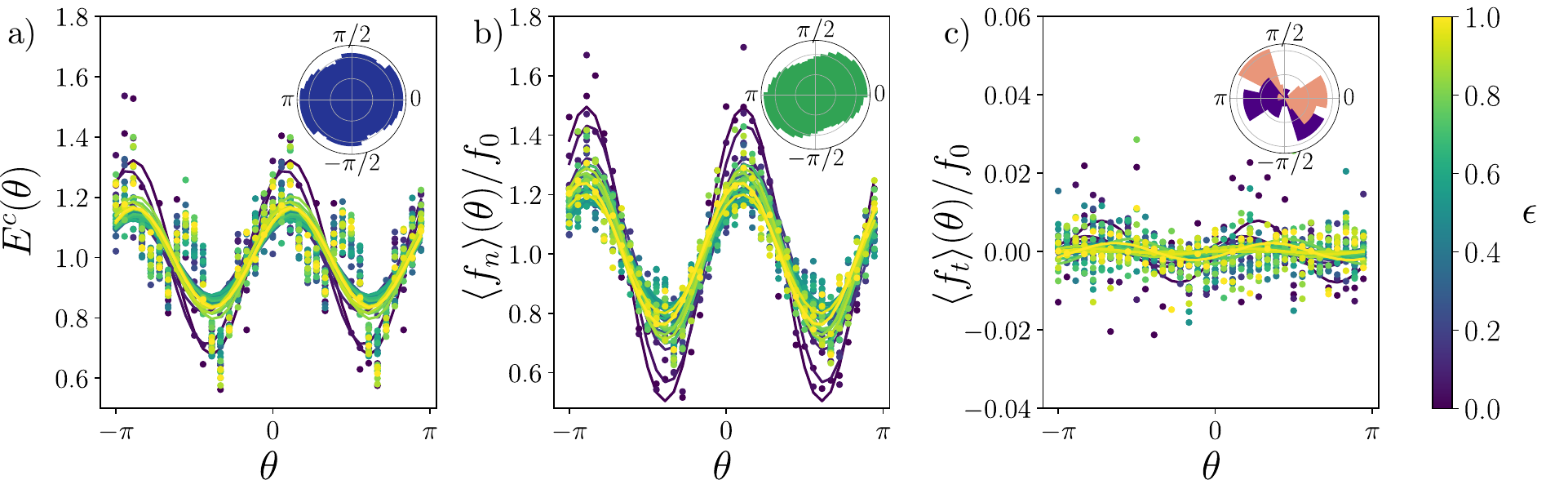}
\caption{\label{fig:angdist} Normalized angular distributions of the (a) contacts $E^c(\theta)$, (b) normal forces $\langle f_n\rangle (\theta)$, and (c) tangential forces $\langle f_t\rangle(\theta)$ at increasing levels of strain $\epsilon$ (dark purple to yellow). The contacts are normalized by the average number of contacts $E_0$, and the forces are normalized by the mean normal force $f_0$ at each given strain. Sample angular histograms of each anisotropic metric are shown in the respective insets, which have been averaged over the strain range shown in the main panel. $\theta = 0$ corresponds to the radial direction, and $\theta = \pi/2$ corresponds to the azimuthal direction. Fits to the respective Fourier expansions are indicated by the solid, sinusoidal line (Eqs.~\ref{eq:expansions}(a)-(c)). The normalized contacts and normalized normal forces show a decrease in anisotropy with increasing strain.}
\end{figure*}

In general, these distributions are approximated by a first-order Fourier expansion such that \cite{rothenburg_analytical_1989,azemaStressstrainBehaviorGeometrical2010,li_stressforcefabric_2013}:
\begin{subequations}
\begin{align}
E^c(\theta) & = \frac{1}{2\pi}\left[1+a \cos 2\left(\theta - \theta_a\right)\right], \label{eq:contacts}\\
\langle f_n \rangle (\theta) & = f_0\left[1+a_n \cos 2\left(\theta - \theta_a\right)\right], \label{eq:normals} \\
\langle f_t \rangle (\theta) & = f_0 a_t \sin 2\left(\theta - \theta_a\right) \label{eq:tangents},
\end{align}
\label{eq:expansions}
\end{subequations}
 where $\theta_a$ is the angle corresponding to the principle stress direction, $f_0$ is the average magnitude of the normal force, and $a, a_n, a_t$ are the amplitudes of the leading order Fourier modes for the contact, normal force and tangential force distributions, respectively.

Examples of the three distributions presented in Eqs~\ref{eq:expansions}(a)-(c) are shown in Fig.~\ref{fig:angdist}
 as polar plots averaged over the entire deformation (inset) and using a colour scale at varying strain indicated from purple (dark) to yellow (light) in increments of 0.04. We plot the change in these anisotropic measures as a function of relative angle between the radial direction and branch vector, where $\theta = 0$ indicates alignment with the radial direction and $\theta = \pi/2$ indicates alignment with the azimuthal direction. We expect the primary direction of the normal forces and contacts to align with the $\theta \sim \pi/4$ direction as the load is distributed through the shear band. This is true for the normal forces, indicating that the majority of the load is transmitted in the shear direction. The contact distribution follows a similar trend, however there is an additional secondary feature that appears as a bump aligned in the azimuthal direction $\theta = \pi/2$. Two possible reasons for the increased number of contacts near the azimuthal direction is either due to an artifact during contact detection, or an indication of a higher order Fourier mode, possibly present due to the constant area boundary condition. Regardless, this feature is small relative to the main signal, and we use only one Fourier mode to describe the data. The tangential distributions show only a weak signal above the noise, indicating that there is no strong preference for average tangential forces to be either clockwise (purple) or counterclockwise (pink) from the contact normal.

Following~\cite{rothenburg_analytical_1989,li_stressforcefabric_2013}, we expand the stress tensor (Eq.~\ref{eq:stressexpand}) with the Fourier approximations (Eqs.~\ref{eq:expansions} (a)-(c)) to compare the particle scale anisotropies to bulk metrics.  In addition, we substitute the number of contacts per unit area as $N_c/S = 2z\nu/(\pi d^2)$, where $z=2N_c/N_p$ is the coordination number, $\nu$ the packing fraction and $d$ the mean grain diameter. The SFF relation states that the stress tensor can be re-written in terms of metrics related to the topology of the contact and forces network as

\begin{multline}
\sigma_{ij} = {\frac{z \nu}{\pi d } f_0} \, \Biggl[\left(1+\frac{aa_n}{2}\right) \delta_{ij}\\
+ \frac{1}{2} \left(a+a_n+a_t\right)\begin{pmatrix}\cos2\theta_a & \sin{2}\theta_a\\\sin 2\theta_a & -\cos{2}\theta_a\end{pmatrix}\Biggr],
\label{eq:anisotropystress}
\end{multline}
Then, inserting Eq.~\ref{eq:anisotropystress} into Eq.~\ref{eq:mu}, and neglecting cross product between the anisotropies, $\mu$ results in a simple sum rule:
\begin{equation}
    \mu = \frac{1}{2}(a+a_n+a_t),
    \label{eq:SFF}
\end{equation}
which directly relates the bulk friction coefficient to the anisotropy amplitude, indicating that the resistance to deformation in a granular material simply depends on the amount of anisotropy that is present in the granular assembly.

\begin{figure}
\includegraphics[width=\columnwidth]{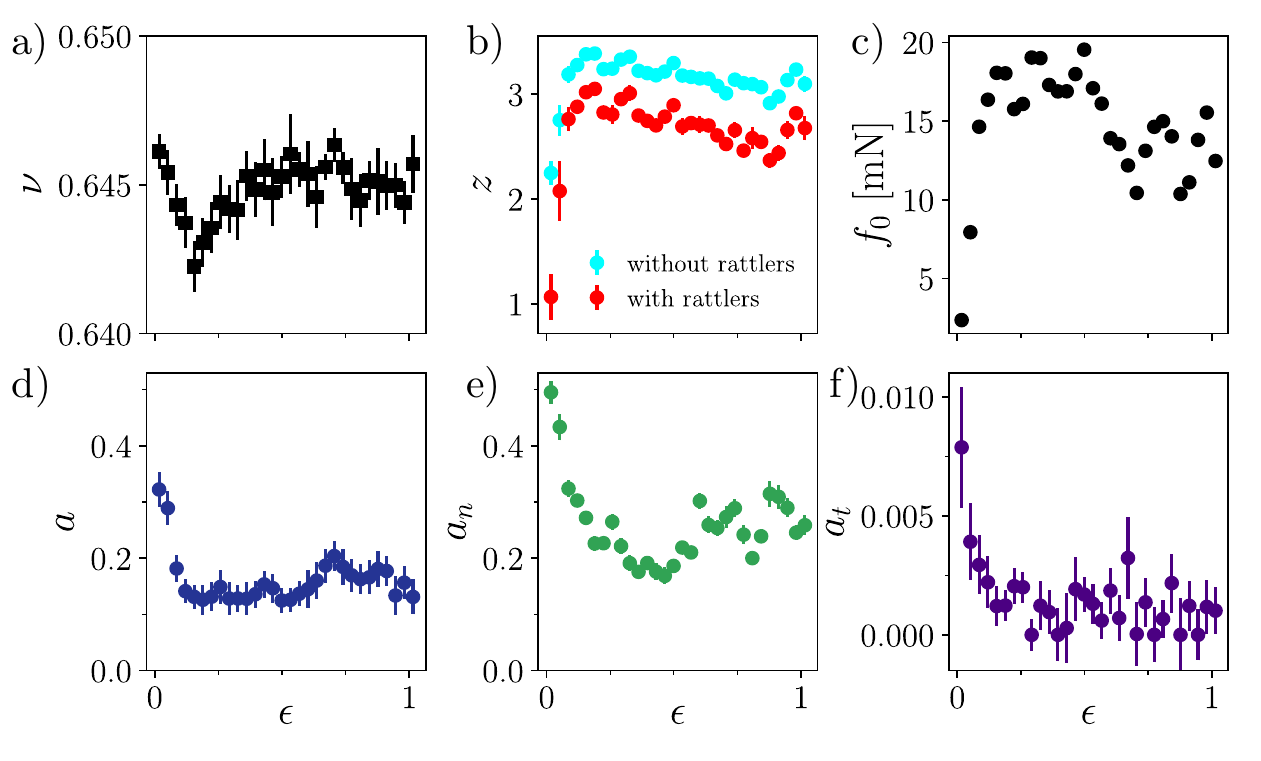}[b]
\caption{Panels (a)-(c) display the evolution of the packing fraction $\nu$, coordination number $z$  with (red) rattlers and without (cyan) rattlers (i.e., particles with less than 2 contacts) and the mean normal force $f_0$, respectively, as a function of strain $\epsilon$. Panels (d)-(f) display the leading term Fourier coefficients representing the angular distribution of contacts ($a$), average normal forces ($a_n$), and tangential forces ($a_t$), respectively, within a strained granular packing under shear. These coefficients quantify the degree of anisotropy observed in each metric within the packing.}
\label{fig:micro}
\end{figure}

To consider the microstructure of the packing, we measure the first order metrics of connectivity. Fig.\ref{fig:micro}(a)-(c) shows the evolution of the packing fraction $\nu$, coordination number $z$ (with or without considering rattlers) and mean normal force $f_0$ as a function of strain. The solid fraction can be considered as constant throughout the strain, while $z$ with (red) or without (cyan) rattlers first increases with strain before saturating around a value of $2$, $3$, respectively. Finally, $f_0$ follows the same trend, first increasing before tending to a constant value in the steady state (also referred as ``critical state'' in soil mechanics).

To estimate the fabric and force networks anisotropies, we fit the angular distribution of contacts, normal forces and tangential forces (Fig.~\ref{fig:angdist}) to the predicted Fourier decomposition in Eqs.~\ref{eq:expansions}(a)-(c). We take the principle stress direction $\theta_a$ and the amplitudes $a$, $a_n$ and $a_t$ to be free fit parameters, where the anisotropy corresponds to the amplitudes. The anisotropies are plotted as a function of strain in Fig.~\ref{fig:micro}(d)-(f); note that during the initial strain there is a strong decrease in value for all three metrics, indicating a decrease in overall anisotropy. At larger strains the three amplitudes tend toward a constant value, meaning that a steady state is reached. In the steady state the relative magnitude of $a$ and $a_n$ are comparable, which is consistent with simulated systems of disks~\cite{azema_shear_2017, rothenburg_analytical_1989}, while $a_t$ is an order of magnitude smaller. This indicates that the tangential forces, and consequently the inter-particle friction, has only a small effect in our system on the bulk friction, due to $a_t$ being smaller than those reported in  simulations~\cite{binaree_combined_2020, azema_shear_2017}.  

\begin{figure}
\includegraphics[width=\columnwidth]{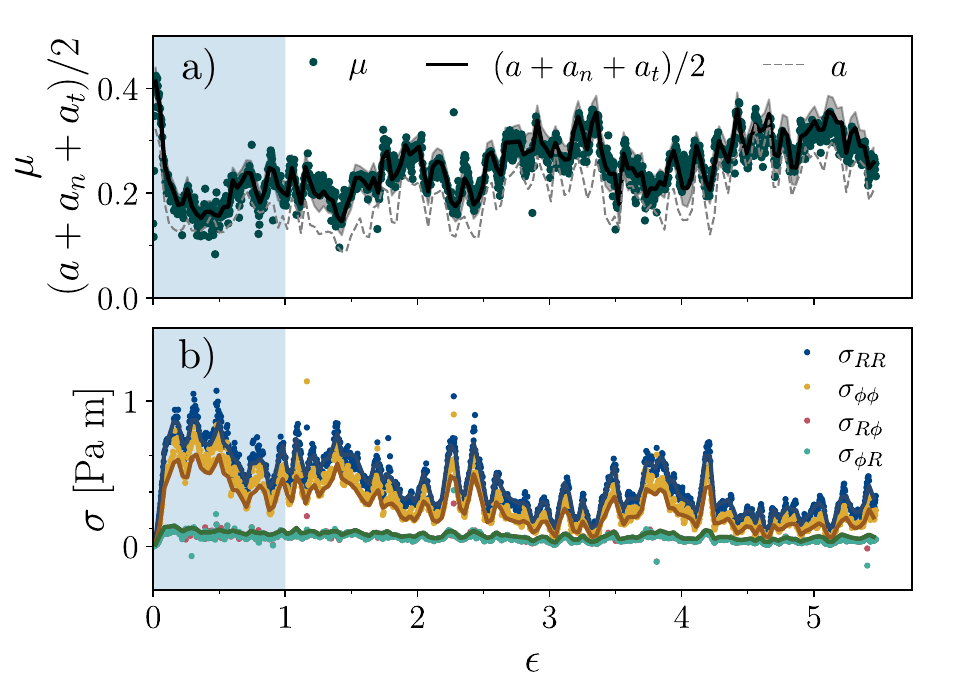}
\caption{ Panel (a) illustrates the bulk friction coefficient of the sheared granular packing as a function of shear, with the summed anisotropies overlaid in black and the standard error in the fit represented as grey filled bars spanning the best fit. The fabric anisotropy $a$ is plotted with a black dashed line for comparison. The blue panel highlights the strain $\epsilon$ depicted in fig.\ref{fig:micro} (a)-(f). Panel (b) presents the four components of the strain tensor, $\sigma$, as measured by Eq.~\ref{eq:stresstensor}, with principal and secondary stress axes oriented radially and azimuthally, respectively. The solid lines represent the stress tensor calculated from the anisotropies using Eq.~\ref{eq:anisotropystress}.}
\label{fig:macromu}
\end{figure}

Figure \ref{fig:macromu}(a) shows the evolution of $\mu$, measured directly from the components of the granular stress tensor (Eq.\ref{eq:mu}) and as predicted by Eq.\ref{eq:SFF}. The prediction from the SFF relation is excellent throughout the deformation and matches the measured value even through large rearrangements and stick slip events which are indicated by large spikes in $\mu$. This result shows for the first time an experimental test of the SFF relationship. For an annular shear cell, we expect $\mu$ to initially decrease as it undergoes strain softening before reaching a steady state~\cite{liu_macroscopic_2020}. Based on the SFF relation, this indicates that the fabric and forces are initially more anisotropic before becoming less anisotropic with increased strain. Notably, $\mu$ is not solely determined by the number of forces (see supplemental video~\cite{video}), as there are strong fluctuations in the number of contact forces for a given value of $\mu$.

To explore the applicability of the SFF relation to systems without contact force data, in Fig.~\ref{fig:macromu}(a) we show the effect of the fabric anisotropy alone by plotting only $a$ as a function of strain (dashed black line). Overall, $a$ follows the overall pattern of $\mu$, including the initial transient and stick-slip events, but underestimates bulk friction by roughly 30\%. Because $a$ and $a_n$ are typically of the same magnitude, while the tangential component is small, using the fabric anisotropy alone provides a good estimate of bulk friction at least for circular particles with low polydispersity \cite{azema_shear_2017,cantorMicrostructuralAnalysisSheared2020}.

As well, we plot the stress tensor components, as a function of strain, in Fig~\ref{fig:macromu}(b). The radial and azimuthal stress increase during strain, and then fluctuate as the particles undergo rearrangements during stick slip events. The off-axis components of the stress tensor remain small. Overlaid on the calculated stress tensor components are the predictions for the stress tensor based on Eq.~\ref{eq:anisotropystress}. Again, these agree well with the measured values. To gain a physical understanding of the influence of the microscale on the stress tensor, we replace the sum of anisotropies by $\mu$ in Eq.\ref{eq:anisotropystress}, and neglect the cross product $a\,a_n$. The rapid increase in $\sigma_{RR}$ and $\sigma_{\phi \phi}$, mainly results from the increases in $z$ and $f_0$. Moreover, in the steady state $z\nu f_0 (1+\mu)/(\pi d)$ provide a fairly good approximation of both $\sigma_{RR}$ and $\sigma_{\phi \phi}$.

In this study, we have shown an experimental validation of the SFF relation using photoelastic particles in an annular shear cell. Success of the SFF relation indicates that the resistance to deformation of a granular material arises due to anisotropies in fabric and forces. Importantly, the SFF relation remains valid even as transient stress builds up and releases through stick slip events, far beyond the static loading for which the framework was initially developed. We note that this framework does not address the role of mesoscale structure and therefore does not encompass prediction of rigidity~\cite{liu_spongelike_2021}; however, the SFF relation is a promising pathway to address the open challenge of connecting fluid-like granular theories~\cite{kamrin_nonlocal_2012, jop_constitutive_2006, berzi_granular_2024} to quasi-static granular behaviour. Finally, the framework may additionally provide a way to link fabric anisotropy to bulk friction in experimental systems for which vector contact forces cannot be measured, as illustrated by the success of the contact anisotropy $a$ alone. We note that in the absence of force information, the geometric anisotropy provides a good estimate of bulk friction,at least when  slightly polydisperse discs are considered \cite{azema_shear_2017,binaree_combined_2020}, and could be easily applied to other deformable assemblies 
\cite{cardenasExperimentalValidationMicromechanically2022}, including cellular tissues, foams, and dense colloid suspensions.

\begin{acknowledgments}
{\bf Acknowledgments:} The authors thank Jonathan Bar\'es and Tejas Murthy for fruitful discussions, and Ben McMillan for improvements to the PEGS photoelastic solver. The authors thank the Lorentz Center for hosting the workshop Getting into Shape where the idea for this paper was formed. C.L.L. acknowledges funding from the NSERC postdoctoral fellowship, and K.E.D and C.L.L from the National Science Foundation (DMR-2104986).
\end{acknowledgments}


\begin{thebibliography}{42}%
\makeatletter
\providecommand \@ifxundefined [1]{%
 \@ifx{#1\undefined}
}%
\providecommand \@ifnum [1]{%
 \ifnum #1\expandafter \@firstoftwo
 \else \expandafter \@secondoftwo
 \fi
}%
\providecommand \@ifx [1]{%
 \ifx #1\expandafter \@firstoftwo
 \else \expandafter \@secondoftwo
 \fi
}%
\providecommand \natexlab [1]{#1}%
\providecommand \enquote  [1]{``#1''}%
\providecommand \bibnamefont  [1]{#1}%
\providecommand \bibfnamefont [1]{#1}%
\providecommand \citenamefont [1]{#1}%
\providecommand \href@noop [0]{\@secondoftwo}%
\providecommand \href [0]{\begingroup \@sanitize@url \@href}%
\providecommand \@href[1]{\@@startlink{#1}\@@href}%
\providecommand \@@href[1]{\endgroup#1\@@endlink}%
\providecommand \@sanitize@url [0]{\catcode `\\12\catcode `\$12\catcode
  `\&12\catcode `\#12\catcode `\^12\catcode `\_12\catcode `\%12\relax}%
\providecommand \@@startlink[1]{}%
\providecommand \@@endlink[0]{}%
\providecommand \url  [0]{\begingroup\@sanitize@url \@url }%
\providecommand \@url [1]{\endgroup\@href {#1}{\urlprefix }}%
\providecommand \urlprefix  [0]{URL }%
\providecommand \Eprint [0]{\href }%
\providecommand \doibase [0]{https://doi.org/}%
\providecommand \selectlanguage [0]{\@gobble}%
\providecommand \bibinfo  [0]{\@secondoftwo}%
\providecommand \bibfield  [0]{\@secondoftwo}%
\providecommand \translation [1]{[#1]}%
\providecommand \BibitemOpen [0]{}%
\providecommand \bibitemStop [0]{}%
\providecommand \bibitemNoStop [0]{.\EOS\space}%
\providecommand \EOS [0]{\spacefactor3000\relax}%
\providecommand \BibitemShut  [1]{\csname bibitem#1\endcsname}%
\let\auto@bib@innerbib\@empty
\bibitem [{\citenamefont {Tordesillas}\ \emph {et~al.}(2004)\citenamefont
  {Tordesillas}, \citenamefont {Walsh},\ and\ \citenamefont
  {Gardiner}}]{tordesillas_bridging_2004}%
  \BibitemOpen
  \bibfield  {author} {\bibinfo {author} {\bibfnamefont {A.}~\bibnamefont
  {Tordesillas}}, \bibinfo {author} {\bibfnamefont {S.~D.~C.}\ \bibnamefont
  {Walsh}},\ and\ \bibinfo {author} {\bibfnamefont {B.~S.}\ \bibnamefont
  {Gardiner}},\ }\bibfield  {title} {\bibinfo {title} {Bridging the {Length}
  {Scales}: {Micromechanics} of {Granular} {Media}},\ }\href
  {https://doi.org/10.1023/B:BITN.0000046817.60322.ed} {\bibfield  {journal}
  {\bibinfo  {journal} {BIT Numerical Mathematics}\ }\textbf {\bibinfo {volume}
  {44}},\ \bibinfo {pages} {539} (\bibinfo {year} {2004})}\BibitemShut
  {NoStop}%
\bibitem [{\citenamefont {Keys}\ \emph {et~al.}(2007)\citenamefont {Keys},
  \citenamefont {Abate}, \citenamefont {Glotzer},\ and\ \citenamefont
  {Durian}}]{keys_measurement_2007}%
  \BibitemOpen
  \bibfield  {author} {\bibinfo {author} {\bibfnamefont {A.~S.}\ \bibnamefont
  {Keys}}, \bibinfo {author} {\bibfnamefont {A.~R.}\ \bibnamefont {Abate}},
  \bibinfo {author} {\bibfnamefont {S.~C.}\ \bibnamefont {Glotzer}},\ and\
  \bibinfo {author} {\bibfnamefont {D.~J.}\ \bibnamefont {Durian}},\ }\bibfield
   {title} {\bibinfo {title} {Measurement of growing dynamical length scales
  and prediction of the jamming transition in a granular material},\ }\href
  {https://doi.org/10.1038/nphys572} {\bibfield  {journal} {\bibinfo  {journal}
  {Nature Physics}\ }\textbf {\bibinfo {volume} {3}},\ \bibinfo {pages} {260}
  (\bibinfo {year} {2007})}\BibitemShut {NoStop}%
\bibitem [{\citenamefont {Shahin}\ \emph {et~al.}(2022)\citenamefont {Shahin},
  \citenamefont {Herbold}, \citenamefont {Hall},\ and\ \citenamefont
  {Hurley}}]{shahin_quantifying_2022}%
  \BibitemOpen
  \bibfield  {author} {\bibinfo {author} {\bibfnamefont {G.}~\bibnamefont
  {Shahin}}, \bibinfo {author} {\bibfnamefont {E.~B.}\ \bibnamefont {Herbold}},
  \bibinfo {author} {\bibfnamefont {S.~A.}\ \bibnamefont {Hall}},\ and\
  \bibinfo {author} {\bibfnamefont {R.~C.}\ \bibnamefont {Hurley}},\ }\bibfield
   {title} {\bibinfo {title} {Quantifying the hierarchy of structural and
  mechanical length scales in granular systems},\ }\href
  {https://doi.org/10.1016/j.eml.2021.101590} {\bibfield  {journal} {\bibinfo
  {journal} {Extreme Mechanics Letters}\ }\textbf {\bibinfo {volume} {51}},\
  \bibinfo {pages} {101590} (\bibinfo {year} {2022})}\BibitemShut {NoStop}%
\bibitem [{\citenamefont {Goldhirsch}(1999)}]{goldhirsch_scales_1999}%
  \BibitemOpen
  \bibfield  {author} {\bibinfo {author} {\bibfnamefont {I.}~\bibnamefont
  {Goldhirsch}},\ }\bibfield  {title} {\bibinfo {title} {Scales and kinetics of
  granular flows},\ }\href {https://doi.org/10.1063/1.166440} {\bibfield
  {journal} {\bibinfo  {journal} {Chaos: An Interdisciplinary Journal of
  Nonlinear Science}\ }\textbf {\bibinfo {volume} {9}},\ \bibinfo {pages} {659}
  (\bibinfo {year} {1999})}\BibitemShut {NoStop}%
\bibitem [{\citenamefont {Liu}\ \emph {et~al.}(2021)\citenamefont {Liu},
  \citenamefont {Kollmer}, \citenamefont {Daniels}, \citenamefont {Schwarz},\
  and\ \citenamefont {Henkes}}]{liu_spongelike_2021}%
  \BibitemOpen
  \bibfield  {author} {\bibinfo {author} {\bibfnamefont {K.}~\bibnamefont
  {Liu}}, \bibinfo {author} {\bibfnamefont {J.~E.}\ \bibnamefont {Kollmer}},
  \bibinfo {author} {\bibfnamefont {K.~E.}\ \bibnamefont {Daniels}}, \bibinfo
  {author} {\bibfnamefont {J.}~\bibnamefont {Schwarz}},\ and\ \bibinfo {author}
  {\bibfnamefont {S.}~\bibnamefont {Henkes}},\ }\bibfield  {title} {\bibinfo
  {title} {Spongelike {Rigid} {Structures} in {Frictional} {Granular}
  {Packings}},\ }\href {https://doi.org/10.1103/PhysRevLett.126.088002}
  {\bibfield  {journal} {\bibinfo  {journal} {Physical Review Letters}\
  }\textbf {\bibinfo {volume} {126}},\ \bibinfo {pages} {088002} (\bibinfo
  {year} {2021})}\BibitemShut {NoStop}%
\bibitem [{\citenamefont {Radjai}\ \emph {et~al.}(2017)\citenamefont {Radjai},
  \citenamefont {Roux},\ and\ \citenamefont {Daouadji}}]{radjai_modeling_2017}%
  \BibitemOpen
  \bibfield  {author} {\bibinfo {author} {\bibfnamefont {F.}~\bibnamefont
  {Radjai}}, \bibinfo {author} {\bibfnamefont {J.-N.}\ \bibnamefont {Roux}},\
  and\ \bibinfo {author} {\bibfnamefont {A.}~\bibnamefont {Daouadji}},\
  }\bibfield  {title} {\bibinfo {title} {Modeling {Granular} {Materials}:
  {Century}-{Long} {Research} across {Scales}},\ }\href
  {https://doi.org/10.1061/(ASCE)EM.1943-7889.0001196} {\bibfield  {journal}
  {\bibinfo  {journal} {Journal of Engineering Mechanics}\ }\textbf {\bibinfo
  {volume} {143}},\ \bibinfo {pages} {04017002} (\bibinfo {year}
  {2017})}\BibitemShut {NoStop}%
\bibitem [{\citenamefont {Börzsönyi}\ and\ \citenamefont
  {Stannarius}(2013)}]{borzsonyi_granular_2013}%
  \BibitemOpen
  \bibfield  {author} {\bibinfo {author} {\bibfnamefont {T.}~\bibnamefont
  {Börzsönyi}}\ and\ \bibinfo {author} {\bibfnamefont {R.}~\bibnamefont
  {Stannarius}},\ }\bibfield  {title} {\bibinfo {title} {Granular materials
  composed of shape-anisotropic grains},\ }\href
  {https://doi.org/10.1039/C3SM50298H} {\bibfield  {journal} {\bibinfo
  {journal} {Soft Matter}\ }\textbf {\bibinfo {volume} {9}},\ \bibinfo {pages}
  {7401} (\bibinfo {year} {2013})}\BibitemShut {NoStop}%
\bibitem [{\citenamefont {Li}\ and\ \citenamefont {Yu}(2014)}]{li_fabric_2014}%
  \BibitemOpen
  \bibfield  {author} {\bibinfo {author} {\bibfnamefont {X.}~\bibnamefont
  {Li}}\ and\ \bibinfo {author} {\bibfnamefont {H.-S.}\ \bibnamefont {Yu}},\
  }\bibfield  {title} {\bibinfo {title} {Fabric, force and strength
  anisotropies in granular materials: a micromechanical insight},\ }\href
  {https://doi.org/10.1007/s00707-014-1120-6} {\bibfield  {journal} {\bibinfo
  {journal} {Acta Mechanica}\ }\textbf {\bibinfo {volume} {225}},\ \bibinfo
  {pages} {2345} (\bibinfo {year} {2014})}\BibitemShut {NoStop}%
\bibitem [{\citenamefont {Dagois-Bohy}\ \emph {et~al.}(2012)\citenamefont
  {Dagois-Bohy}, \citenamefont {Tighe}, \citenamefont {Simon}, \citenamefont
  {Henkes},\ and\ \citenamefont {van Hecke}}]{dagois-bohy_soft-sphere_2012}%
  \BibitemOpen
  \bibfield  {author} {\bibinfo {author} {\bibfnamefont {S.}~\bibnamefont
  {Dagois-Bohy}}, \bibinfo {author} {\bibfnamefont {B.~P.}\ \bibnamefont
  {Tighe}}, \bibinfo {author} {\bibfnamefont {J.}~\bibnamefont {Simon}},
  \bibinfo {author} {\bibfnamefont {S.}~\bibnamefont {Henkes}},\ and\ \bibinfo
  {author} {\bibfnamefont {M.}~\bibnamefont {van Hecke}},\ }\bibfield  {title}
  {\bibinfo {title} {Soft-{Sphere} {Packings} at {Finite} {Pressure} but
  {Unstable} to {Shear}},\ }\href
  {http://prl.aps.org/abstract/PRL/v109/i9/e095703} {\bibfield  {journal}
  {\bibinfo  {journal} {Physical Review Letters}\ }\textbf {\bibinfo {volume}
  {109}} (\bibinfo {year} {2012})}\BibitemShut {NoStop}%
\bibitem [{\citenamefont {Tighe}\ \emph {et~al.}(2010)\citenamefont {Tighe},
  \citenamefont {Snoeijer}, \citenamefont {Vlugt},\ and\ \citenamefont
  {Hecke}}]{tighe_force_2010}%
  \BibitemOpen
  \bibfield  {author} {\bibinfo {author} {\bibfnamefont {B.}~\bibnamefont
  {Tighe}}, \bibinfo {author} {\bibfnamefont {J.}~\bibnamefont {Snoeijer}},
  \bibinfo {author} {\bibfnamefont {T.~J.}\ \bibnamefont {Vlugt}},\ and\
  \bibinfo {author} {\bibfnamefont {M.~v.}\ \bibnamefont {Hecke}},\ }\bibfield
  {title} {\bibinfo {title} {The force network ensemble for granular
  packings},\ }\href {https://doi.org/10.1039/B926592A} {\bibfield  {journal}
  {\bibinfo  {journal} {Soft Matter}\ }\textbf {\bibinfo {volume} {6}},\
  \bibinfo {pages} {2908} (\bibinfo {year} {2010})}\BibitemShut {NoStop}%
\bibitem [{\citenamefont {da~Cruz}\ \emph {et~al.}(2005)\citenamefont
  {da~Cruz}, \citenamefont {Emam}, \citenamefont {Prochnow}, \citenamefont
  {Roux},\ and\ \citenamefont {Chevoir}}]{da_cruz_rheophysics_2005}%
  \BibitemOpen
  \bibfield  {author} {\bibinfo {author} {\bibfnamefont {F.}~\bibnamefont
  {da~Cruz}}, \bibinfo {author} {\bibfnamefont {S.}~\bibnamefont {Emam}},
  \bibinfo {author} {\bibfnamefont {M.}~\bibnamefont {Prochnow}}, \bibinfo
  {author} {\bibfnamefont {J.-N.}\ \bibnamefont {Roux}},\ and\ \bibinfo
  {author} {\bibfnamefont {F.}~\bibnamefont {Chevoir}},\ }\bibfield  {title}
  {\bibinfo {title} {Rheophysics of dense granular materials: {Discrete}
  simulation of plane shear flows},\ }\href
  {https://doi.org/10.1103/PhysRevE.72.021309} {\bibfield  {journal} {\bibinfo
  {journal} {Physical Review E}\ }\textbf {\bibinfo {volume} {72}},\ \bibinfo
  {pages} {021309} (\bibinfo {year} {2005})}\BibitemShut {NoStop}%
\bibitem [{\citenamefont {Az{\'e}ma}\ and\ \citenamefont
  {Radja{\"i}}(2014)}]{azemaInternalStructureInertial2014}%
  \BibitemOpen
  \bibfield  {author} {\bibinfo {author} {\bibfnamefont {E.}~\bibnamefont
  {Az{\'e}ma}}\ and\ \bibinfo {author} {\bibfnamefont {F.}~\bibnamefont
  {Radja{\"i}}},\ }\bibfield  {title} {\bibinfo {title} {Internal {{Structure}}
  of {{Inertial Granular Flows}}},\ }\bibfield  {journal} {\bibinfo  {journal}
  {Physical Review Letters}\ }\textbf {\bibinfo {volume} {112}},\ \href
  {https://doi.org/10.1103/PhysRevLett.112.078001}
  {10.1103/PhysRevLett.112.078001} (\bibinfo {year} {2014})\BibitemShut
  {NoStop}%
\bibitem [{\citenamefont {Jop}\ \emph {et~al.}(2006)\citenamefont {Jop},
  \citenamefont {Forterre},\ and\ \citenamefont
  {Pouliquen}}]{jop_constitutive_2006}%
  \BibitemOpen
  \bibfield  {author} {\bibinfo {author} {\bibfnamefont {P.}~\bibnamefont
  {Jop}}, \bibinfo {author} {\bibfnamefont {Y.}~\bibnamefont {Forterre}},\ and\
  \bibinfo {author} {\bibfnamefont {O.}~\bibnamefont {Pouliquen}},\ }\bibfield
  {title} {\bibinfo {title} {A constitutive law for dense granular flows},\
  }\href {https://doi.org/10.1038/nature04801} {\bibfield  {journal} {\bibinfo
  {journal} {Nature}\ }\textbf {\bibinfo {volume} {441}},\ \bibinfo {pages}
  {727} (\bibinfo {year} {2006})},\ \bibinfo {note} {publisher: Nature
  Publishing Group}\BibitemShut {NoStop}%
\bibitem [{\citenamefont {Kamrin}\ and\ \citenamefont
  {Koval}(2012)}]{kamrin_nonlocal_2012}%
  \BibitemOpen
  \bibfield  {author} {\bibinfo {author} {\bibfnamefont {K.}~\bibnamefont
  {Kamrin}}\ and\ \bibinfo {author} {\bibfnamefont {G.}~\bibnamefont {Koval}},\
  }\bibfield  {title} {\bibinfo {title} {Nonlocal {Constitutive} {Relation} for
  {Steady} {Granular} {Flow}},\ }\href
  {https://doi.org/10.1103/PhysRevLett.108.178301} {\bibfield  {journal}
  {\bibinfo  {journal} {Physical Review Letters}\ }\textbf {\bibinfo {volume}
  {108}},\ \bibinfo {pages} {178301} (\bibinfo {year} {2012})}\BibitemShut
  {NoStop}%
\bibitem [{\citenamefont {Berzi}(2024)}]{berzi_granular_2024}%
  \BibitemOpen
  \bibfield  {author} {\bibinfo {author} {\bibfnamefont {D.}~\bibnamefont
  {Berzi}},\ }\bibfield  {title} {\bibinfo {title} {On granular flows: {From}
  kinetic theory to inertial rheology and nonlocal constitutive models},\
  }\href {https://doi.org/10.1103/PhysRevFluids.9.034304} {\bibfield  {journal}
  {\bibinfo  {journal} {Physical Review Fluids}\ }\textbf {\bibinfo {volume}
  {9}},\ \bibinfo {pages} {034304} (\bibinfo {year} {2024})}\BibitemShut
  {NoStop}%
\bibitem [{\citenamefont {Cates}\ \emph {et~al.}(1998)\citenamefont {Cates},
  \citenamefont {Wittmer}, \citenamefont {Bouchaud},\ and\ \citenamefont
  {Claudin}}]{cates_jamming_1998}%
  \BibitemOpen
  \bibfield  {author} {\bibinfo {author} {\bibfnamefont {M.~E.}\ \bibnamefont
  {Cates}}, \bibinfo {author} {\bibfnamefont {J.~P.}\ \bibnamefont {Wittmer}},
  \bibinfo {author} {\bibfnamefont {J.-P.}\ \bibnamefont {Bouchaud}},\ and\
  \bibinfo {author} {\bibfnamefont {P.}~\bibnamefont {Claudin}},\ }\bibfield
  {title} {\bibinfo {title} {Jamming, {Force} {Chains}, and {Fragile}
  {Matter}},\ }\href {https://doi.org/10.1103/PhysRevLett.81.1841} {\bibfield
  {journal} {\bibinfo  {journal} {Physical Review Letters}\ }\textbf {\bibinfo
  {volume} {81}},\ \bibinfo {pages} {1841} (\bibinfo {year}
  {1998})}\BibitemShut {NoStop}%
\bibitem [{\citenamefont {Chang}\ \emph {et~al.}(2011)\citenamefont {Chang},
  \citenamefont {Yin},\ and\ \citenamefont
  {Hicher}}]{chang_micromechanical_2011}%
  \BibitemOpen
  \bibfield  {author} {\bibinfo {author} {\bibfnamefont {C.~S.}\ \bibnamefont
  {Chang}}, \bibinfo {author} {\bibfnamefont {Z.-Y.}\ \bibnamefont {Yin}},\
  and\ \bibinfo {author} {\bibfnamefont {P.-Y.}\ \bibnamefont {Hicher}},\
  }\bibfield  {title} {\bibinfo {title} {Micromechanical {Analysis} for
  {Interparticle} and {Assembly} {Instability} of {Sand}},\ }\href
  {https://doi.org/10.1061/(ASCE)EM.1943-7889.0000204} {\bibfield  {journal}
  {\bibinfo  {journal} {Journal of Engineering Mechanics}\ }\textbf {\bibinfo
  {volume} {137}},\ \bibinfo {pages} {155} (\bibinfo {year}
  {2011})}\BibitemShut {NoStop}%
\bibitem [{\citenamefont {Majmudar}\ and\ \citenamefont
  {Behringer}(2005)}]{majmudar_contact_2005}%
  \BibitemOpen
  \bibfield  {author} {\bibinfo {author} {\bibfnamefont {T.~S.}\ \bibnamefont
  {Majmudar}}\ and\ \bibinfo {author} {\bibfnamefont {R.~P.}\ \bibnamefont
  {Behringer}},\ }\bibfield  {title} {\bibinfo {title} {Contact force
  measurements and stress-induced anisotropy in granular materials},\ }\href
  {https://doi.org/10.1038/nature03805} {\bibfield  {journal} {\bibinfo
  {journal} {Nature}\ }\textbf {\bibinfo {volume} {435}},\ \bibinfo {pages}
  {1079} (\bibinfo {year} {2005})}\BibitemShut {NoStop}%
\bibitem [{\citenamefont {Wang}\ \emph {et~al.}(2017)\citenamefont {Wang},
  \citenamefont {Fu}, \citenamefont {Tong}, \citenamefont {Zhang},\ and\
  \citenamefont {Dafalias}}]{wang_strength_2017}%
  \BibitemOpen
  \bibfield  {author} {\bibinfo {author} {\bibfnamefont {R.}~\bibnamefont
  {Wang}}, \bibinfo {author} {\bibfnamefont {P.}~\bibnamefont {Fu}}, \bibinfo
  {author} {\bibfnamefont {Z.}~\bibnamefont {Tong}}, \bibinfo {author}
  {\bibfnamefont {J.-M.}\ \bibnamefont {Zhang}},\ and\ \bibinfo {author}
  {\bibfnamefont {Y.~F.}\ \bibnamefont {Dafalias}},\ }\bibfield  {title}
  {\bibinfo {title} {Strength anisotropy of granular material consisting of
  perfectly round particles},\ }\href {https://doi.org/10.1002/nag.2699}
  {\bibfield  {journal} {\bibinfo  {journal} {International Journal for
  Numerical and Analytical Methods in Geomechanics}\ }\textbf {\bibinfo
  {volume} {41}},\ \bibinfo {pages} {1758} (\bibinfo {year}
  {2017})}\BibitemShut {NoStop}%
\bibitem [{\citenamefont {Shi}\ and\ \citenamefont
  {Guo}(2018)}]{shi_fabric_2018}%
  \BibitemOpen
  \bibfield  {author} {\bibinfo {author} {\bibfnamefont {J.}~\bibnamefont
  {Shi}}\ and\ \bibinfo {author} {\bibfnamefont {P.}~\bibnamefont {Guo}},\
  }\bibfield  {title} {\bibinfo {title} {Fabric evolution of granular materials
  along imposed stress paths},\ }\href
  {https://doi.org/10.1007/s11440-018-0665-2} {\bibfield  {journal} {\bibinfo
  {journal} {Acta Geotechnica}\ }\textbf {\bibinfo {volume} {13}},\ \bibinfo
  {pages} {1341} (\bibinfo {year} {2018})}\BibitemShut {NoStop}%
\bibitem [{\citenamefont {Hurley}\ \emph {et~al.}(2016)\citenamefont {Hurley},
  \citenamefont {Hall}, \citenamefont {Andrade},\ and\ \citenamefont
  {Wright}}]{hurley_quantifying_2016}%
  \BibitemOpen
  \bibfield  {author} {\bibinfo {author} {\bibfnamefont {R.~C.}\ \bibnamefont
  {Hurley}}, \bibinfo {author} {\bibfnamefont {S.~A.}\ \bibnamefont {Hall}},
  \bibinfo {author} {\bibfnamefont {J.~E.}\ \bibnamefont {Andrade}},\ and\
  \bibinfo {author} {\bibfnamefont {J.}~\bibnamefont {Wright}},\ }\bibfield
  {title} {\bibinfo {title} {Quantifying {Interparticle} {Forces} and
  {Heterogeneity} in {3D} {Granular} {Materials}},\ }\href
  {https://doi.org/10.1103/PhysRevLett.117.098005} {\bibfield  {journal}
  {\bibinfo  {journal} {Physical Review Letters}\ }\textbf {\bibinfo {volume}
  {117}},\ \bibinfo {pages} {098005} (\bibinfo {year} {2016})}\BibitemShut
  {NoStop}%
\bibitem [{\citenamefont {Bi}\ \emph {et~al.}(2011)\citenamefont {Bi},
  \citenamefont {Zhang}, \citenamefont {Chakraborty},\ and\ \citenamefont
  {Behringer}}]{bi_jamming_2011}%
  \BibitemOpen
  \bibfield  {author} {\bibinfo {author} {\bibfnamefont {D.}~\bibnamefont
  {Bi}}, \bibinfo {author} {\bibfnamefont {J.}~\bibnamefont {Zhang}}, \bibinfo
  {author} {\bibfnamefont {B.}~\bibnamefont {Chakraborty}},\ and\ \bibinfo
  {author} {\bibfnamefont {R.~P.}\ \bibnamefont {Behringer}},\ }\bibfield
  {title} {\bibinfo {title} {Jamming by shear},\ }\href
  {https://doi.org/10.1038/nature10667} {\bibfield  {journal} {\bibinfo
  {journal} {Nature}\ }\textbf {\bibinfo {volume} {480}},\ \bibinfo {pages}
  {355} (\bibinfo {year} {2011})}\BibitemShut {NoStop}%
\bibitem [{\citenamefont {Rothenburg}\ and\ \citenamefont
  {Bathurst}(1989)}]{rothenburg_analytical_1989}%
  \BibitemOpen
  \bibfield  {author} {\bibinfo {author} {\bibfnamefont {L.}~\bibnamefont
  {Rothenburg}}\ and\ \bibinfo {author} {\bibfnamefont {R.~J.}\ \bibnamefont
  {Bathurst}},\ }\bibfield  {title} {\bibinfo {title} {Analytical study of
  induced anisotropy in idealized granular materials},\ }\href
  {https://doi.org/10.1680/geot.1989.39.4.601} {\bibfield  {journal} {\bibinfo
  {journal} {Géotechnique}\ }\textbf {\bibinfo {volume} {39}},\ \bibinfo
  {pages} {601} (\bibinfo {year} {1989})}\BibitemShut {NoStop}%
\bibitem [{\citenamefont {Azéma}\ \emph {et~al.}(2007)\citenamefont {Azéma},
  \citenamefont {Radjaï}, \citenamefont {Peyroux},\ and\ \citenamefont
  {Saussine}}]{azema_force_2007}%
  \BibitemOpen
  \bibfield  {author} {\bibinfo {author} {\bibfnamefont {E.}~\bibnamefont
  {Azéma}}, \bibinfo {author} {\bibfnamefont {F.}~\bibnamefont {Radjaï}},
  \bibinfo {author} {\bibfnamefont {R.}~\bibnamefont {Peyroux}},\ and\ \bibinfo
  {author} {\bibfnamefont {G.}~\bibnamefont {Saussine}},\ }\bibfield  {title}
  {\bibinfo {title} {Force transmission in a packing of pentagonal particles},\
  }\href {https://doi.org/10.1103/PhysRevE.76.011301} {\bibfield  {journal}
  {\bibinfo  {journal} {Physical Review E}\ }\textbf {\bibinfo {volume} {76}},\
  \bibinfo {pages} {011301} (\bibinfo {year} {2007})}\BibitemShut {NoStop}%
\bibitem [{\citenamefont {Azéma}\ \emph {et~al.}(2017)\citenamefont {Azéma},
  \citenamefont {Linero}, \citenamefont {Estrada},\ and\ \citenamefont
  {Lizcano}}]{azema_shear_2017}%
  \BibitemOpen
  \bibfield  {author} {\bibinfo {author} {\bibfnamefont {E.}~\bibnamefont
  {Azéma}}, \bibinfo {author} {\bibfnamefont {S.}~\bibnamefont {Linero}},
  \bibinfo {author} {\bibfnamefont {N.}~\bibnamefont {Estrada}},\ and\ \bibinfo
  {author} {\bibfnamefont {A.}~\bibnamefont {Lizcano}},\ }\bibfield  {title}
  {\bibinfo {title} {Shear strength and microstructure of polydisperse
  packings: {The} effect of size span and shape of particle size
  distribution},\ }\href {https://doi.org/10.1103/PhysRevE.96.022902}
  {\bibfield  {journal} {\bibinfo  {journal} {Physical Review E}\ }\textbf
  {\bibinfo {volume} {96}},\ \bibinfo {pages} {022902} (\bibinfo {year}
  {2017})}\BibitemShut {NoStop}%
\bibitem [{\citenamefont {Binaree}\ \emph {et~al.}(2020)\citenamefont
  {Binaree}, \citenamefont {Azéma}, \citenamefont {Estrada}, \citenamefont
  {Renouf},\ and\ \citenamefont {Preechawuttipong}}]{binaree_combined_2020}%
  \BibitemOpen
  \bibfield  {author} {\bibinfo {author} {\bibfnamefont {T.}~\bibnamefont
  {Binaree}}, \bibinfo {author} {\bibfnamefont {E.}~\bibnamefont {Azéma}},
  \bibinfo {author} {\bibfnamefont {N.}~\bibnamefont {Estrada}}, \bibinfo
  {author} {\bibfnamefont {M.}~\bibnamefont {Renouf}},\ and\ \bibinfo {author}
  {\bibfnamefont {I.}~\bibnamefont {Preechawuttipong}},\ }\bibfield  {title}
  {\bibinfo {title} {Combined effects of contact friction and particle shape on
  strength properties and microstructure of sheared granular media},\ }\href
  {https://doi.org/10.1103/PhysRevE.102.022901} {\bibfield  {journal} {\bibinfo
   {journal} {Physical Review E}\ }\textbf {\bibinfo {volume} {102}},\ \bibinfo
  {pages} {022901} (\bibinfo {year} {2020})}\BibitemShut {NoStop}%
\bibitem [{\citenamefont {Zhou}\ and\ \citenamefont
  {Xu}(2024)}]{zhou_microscopic_2024}%
  \BibitemOpen
  \bibfield  {author} {\bibinfo {author} {\bibfnamefont {W.}~\bibnamefont
  {Zhou}}\ and\ \bibinfo {author} {\bibfnamefont {M.}~\bibnamefont {Xu}},\
  }\bibfield  {title} {\bibinfo {title} {Microscopic analysis of the nonlinear
  stiffness of granular materials at small-to-medium strain},\ }\href
  {https://doi.org/10.1016/j.compgeo.2023.105859} {\bibfield  {journal}
  {\bibinfo  {journal} {Computers and Geotechnics}\ }\textbf {\bibinfo {volume}
  {165}},\ \bibinfo {pages} {105859} (\bibinfo {year} {2024})}\BibitemShut
  {NoStop}%
\bibitem [{\citenamefont {Liu}\ \emph {et~al.}(2022)\citenamefont {Liu},
  \citenamefont {Wautier}, \citenamefont {Nicot}, \citenamefont {Darve},\ and\
  \citenamefont {Zhou}}]{liu_how_2022}%
  \BibitemOpen
  \bibfield  {author} {\bibinfo {author} {\bibfnamefont {J.}~\bibnamefont
  {Liu}}, \bibinfo {author} {\bibfnamefont {A.}~\bibnamefont {Wautier}},
  \bibinfo {author} {\bibfnamefont {F.}~\bibnamefont {Nicot}}, \bibinfo
  {author} {\bibfnamefont {F.}~\bibnamefont {Darve}},\ and\ \bibinfo {author}
  {\bibfnamefont {W.}~\bibnamefont {Zhou}},\ }\bibfield  {title} {\bibinfo
  {title} {How meso shear chains bridge multiscale shear behaviors in granular
  materials: {A} preliminary study},\ }\href
  {https://doi.org/10.1016/j.ijsolstr.2022.111835} {\bibfield  {journal}
  {\bibinfo  {journal} {International Journal of Solids and Structures}\
  }\textbf {\bibinfo {volume} {252}},\ \bibinfo {pages} {111835} (\bibinfo
  {year} {2022})}\BibitemShut {NoStop}%
\bibitem [{\citenamefont {Li}\ and\ \citenamefont
  {Yu}(2013)}]{li_stressforcefabric_2013}%
  \BibitemOpen
  \bibfield  {author} {\bibinfo {author} {\bibfnamefont {X.}~\bibnamefont
  {Li}}\ and\ \bibinfo {author} {\bibfnamefont {H.~S.}\ \bibnamefont {Yu}},\
  }\bibfield  {title} {\bibinfo {title} {On the stress–force–fabric
  relationship for granular materials},\ }\href
  {https://doi.org/10.1016/j.ijsolstr.2012.12.023} {\bibfield  {journal}
  {\bibinfo  {journal} {International Journal of Solids and Structures}\
  }\textbf {\bibinfo {volume} {50}},\ \bibinfo {pages} {1285} (\bibinfo {year}
  {2013})}\BibitemShut {NoStop}%
\bibitem [{\citenamefont {Abed~Zadeh}\ \emph {et~al.}(2019)\citenamefont
  {Abed~Zadeh}, \citenamefont {Bares}, \citenamefont {Brzinski}, \citenamefont
  {Daniels}, \citenamefont {Dijksman}, \citenamefont {Docquier}, \citenamefont
  {Everitt}, \citenamefont {Kollmer}, \citenamefont {Lantsoght}, \citenamefont
  {Wang}, \citenamefont {Workamp}, \citenamefont {Zhao},\ and\ \citenamefont
  {Zheng}}]{abed_zadeh_enlightening_2019}%
  \BibitemOpen
  \bibfield  {author} {\bibinfo {author} {\bibfnamefont {A.}~\bibnamefont
  {Abed~Zadeh}}, \bibinfo {author} {\bibfnamefont {J.}~\bibnamefont {Bares}},
  \bibinfo {author} {\bibfnamefont {T.~A.}\ \bibnamefont {Brzinski}}, \bibinfo
  {author} {\bibfnamefont {K.~E.}\ \bibnamefont {Daniels}}, \bibinfo {author}
  {\bibfnamefont {J.}~\bibnamefont {Dijksman}}, \bibinfo {author}
  {\bibfnamefont {N.}~\bibnamefont {Docquier}}, \bibinfo {author}
  {\bibfnamefont {H.~O.}\ \bibnamefont {Everitt}}, \bibinfo {author}
  {\bibfnamefont {J.~E.}\ \bibnamefont {Kollmer}}, \bibinfo {author}
  {\bibfnamefont {O.}~\bibnamefont {Lantsoght}}, \bibinfo {author}
  {\bibfnamefont {D.}~\bibnamefont {Wang}}, \bibinfo {author} {\bibfnamefont
  {M.}~\bibnamefont {Workamp}}, \bibinfo {author} {\bibfnamefont
  {Y.}~\bibnamefont {Zhao}},\ and\ \bibinfo {author} {\bibfnamefont
  {H.}~\bibnamefont {Zheng}},\ }\bibfield  {title} {\bibinfo {title}
  {Enlightening force chains: a review of photoelasticimetry in granular
  matter},\ }\href {https://doi.org/10.1007/s10035-019-0942-2} {\bibfield
  {journal} {\bibinfo  {journal} {Granular Matter}\ }\textbf {\bibinfo {volume}
  {21}},\ \bibinfo {pages} {83} (\bibinfo {year} {2019})}\BibitemShut {NoStop}%
\bibitem [{\citenamefont {Daniels}\ \emph {et~al.}(2017)\citenamefont
  {Daniels}, \citenamefont {Kollmer},\ and\ \citenamefont
  {Puckett}}]{daniels_photoelastic_2017}%
  \BibitemOpen
  \bibfield  {author} {\bibinfo {author} {\bibfnamefont {K.~E.}\ \bibnamefont
  {Daniels}}, \bibinfo {author} {\bibfnamefont {J.~E.}\ \bibnamefont
  {Kollmer}},\ and\ \bibinfo {author} {\bibfnamefont {J.~G.}\ \bibnamefont
  {Puckett}},\ }\bibfield  {title} {\bibinfo {title} {Photoelastic force
  measurements in granular materials},\ }\href
  {https://doi.org/10.1063/1.4983049} {\bibfield  {journal} {\bibinfo
  {journal} {Review of Scientific Instruments}\ }\textbf {\bibinfo {volume}
  {88}},\ \bibinfo {pages} {051808} (\bibinfo {year} {2017})}\BibitemShut
  {NoStop}%
\bibitem [{\citenamefont {Kollmer}(2024)}]{kollmer_photo-elastic_nodate}%
  \BibitemOpen
  \bibfield  {author} {\bibinfo {author} {\bibfnamefont {J.~E.}\ \bibnamefont
  {Kollmer}},\ }\href@noop {} {\bibinfo {title} {Photo-{Elastic} {Granular}
  {Solver} ({PEGS}) https://github.com/jekollmer/pegs}} \BibitemShut {NoStop}%
\bibitem [{\citenamefont {Lee}\ \emph {et~al.}(2024)\citenamefont {Lee},
  \citenamefont {Bililign}, \citenamefont {Azéma},\ and\ \citenamefont
  {Daniels}}]{lee_preprint2}%
  \BibitemOpen
  \bibfield  {author} {\bibinfo {author} {\bibfnamefont {C.~L.}\ \bibnamefont
  {Lee}}, \bibinfo {author} {\bibfnamefont {E.}~\bibnamefont {Bililign}},
  \bibinfo {author} {\bibfnamefont {E.}~\bibnamefont {Azéma}},\ and\ \bibinfo
  {author} {\bibfnamefont {K.~E.}\ \bibnamefont {Daniels}},\ }\href
  {https://doi.org/10.48550/ARXIV.2409.08140} {\bibinfo {title}
  {Loading-dependent microscale measures control bulk properties in granular
  material: an experimental test of the stress-force-fabric relation}}
  (\bibinfo {year} {2024})\BibitemShut {NoStop}%
\bibitem [{\citenamefont {Brzinski}\ and\ \citenamefont
  {Daniels}(2018)}]{brzinski_sounds_2018}%
  \BibitemOpen
  \bibfield  {author} {\bibinfo {author} {\bibfnamefont {T.~A.}\ \bibnamefont
  {Brzinski}}\ and\ \bibinfo {author} {\bibfnamefont {K.~E.}\ \bibnamefont
  {Daniels}},\ }\bibfield  {title} {\bibinfo {title} {Sounds of {Failure}:
  {Passive} {Acoustic} {Measurements} of {Excited} {Vibrational} {Modes}},\
  }\href {https://doi.org/10.1103/PhysRevLett.120.218003} {\bibfield  {journal}
  {\bibinfo  {journal} {Physical Review Letters}\ }\textbf {\bibinfo {volume}
  {120}},\ \bibinfo {pages} {218003} (\bibinfo {year} {2018})}\BibitemShut
  {NoStop}%
\bibitem [{\citenamefont {Fazelpour}\ and\ \citenamefont
  {Daniels}(2023)}]{fazelpour_controlling_2023}%
  \BibitemOpen
  \bibfield  {author} {\bibinfo {author} {\bibfnamefont {F.}~\bibnamefont
  {Fazelpour}}\ and\ \bibinfo {author} {\bibfnamefont {K.~E.}\ \bibnamefont
  {Daniels}},\ }\bibfield  {title} {\bibinfo {title} {Controlling rheology via
  boundary conditions in dense granular flows},\ }\href
  {https://doi.org/10.1039/D2SM00683A} {\bibfield  {journal} {\bibinfo
  {journal} {Soft Matter}\ }\textbf {\bibinfo {volume} {19}},\ \bibinfo {pages}
  {2168} (\bibinfo {year} {2023})}\BibitemShut {NoStop}%
\bibitem [{\citenamefont {Kruyt}\ and\ \citenamefont
  {Rothenburg}(2014)}]{kruyt_micromechanical_2014}%
  \BibitemOpen
  \bibfield  {author} {\bibinfo {author} {\bibfnamefont {N.~P.}\ \bibnamefont
  {Kruyt}}\ and\ \bibinfo {author} {\bibfnamefont {L.}~\bibnamefont
  {Rothenburg}},\ }\bibfield  {title} {\bibinfo {title} {On micromechanical
  characteristics of the critical state of two-dimensional granular
  materials},\ }\href {https://doi.org/10.1007/s00707-014-1128-y} {\bibfield
  {journal} {\bibinfo  {journal} {Acta Mechanica}\ }\textbf {\bibinfo {volume}
  {225}},\ \bibinfo {pages} {2301} (\bibinfo {year} {2014})}\BibitemShut
  {NoStop}%
\bibitem [{\citenamefont {Bagi}(1999)}]{bagi_microstructural_1999}%
  \BibitemOpen
  \bibfield  {author} {\bibinfo {author} {\bibfnamefont {K.}~\bibnamefont
  {Bagi}},\ }\bibfield  {title} {\bibinfo {title} {Microstructural {Stress}
  {Tensor} of {Granular} {Assemblies} {With} {Volume} {Forces}},\ }\href
  {https://doi.org/10.1115/1.2791800} {\bibfield  {journal} {\bibinfo
  {journal} {Journal of Applied Mechanics}\ }\textbf {\bibinfo {volume} {66}},\
  \bibinfo {pages} {934} (\bibinfo {year} {1999})}\BibitemShut {NoStop}%
\bibitem [{\citenamefont {Az{\'e}ma}\ and\ \citenamefont
  {Radja{\"i}}(2010)}]{azemaStressstrainBehaviorGeometrical2010}%
  \BibitemOpen
  \bibfield  {author} {\bibinfo {author} {\bibfnamefont {E.}~\bibnamefont
  {Az{\'e}ma}}\ and\ \bibinfo {author} {\bibfnamefont {F.}~\bibnamefont
  {Radja{\"i}}},\ }\bibfield  {title} {\bibinfo {title} {Stress-strain behavior
  and geometrical properties of packings of elongated particles},\ }\href
  {https://doi.org/10.1103/PhysRevE.81.051304} {\bibfield  {journal} {\bibinfo
  {journal} {Physical Review E}\ }\textbf {\bibinfo {volume} {81}},\ \bibinfo
  {pages} {051304} (\bibinfo {year} {2010})}\BibitemShut {NoStop}%
\bibitem [{\citenamefont {Liu}\ \emph {et~al.}(2020)\citenamefont {Liu},
  \citenamefont {Wautier}, \citenamefont {Bonelli}, \citenamefont {Nicot},\
  and\ \citenamefont {Darve}}]{liu_macroscopic_2020}%
  \BibitemOpen
  \bibfield  {author} {\bibinfo {author} {\bibfnamefont {J.}~\bibnamefont
  {Liu}}, \bibinfo {author} {\bibfnamefont {A.}~\bibnamefont {Wautier}},
  \bibinfo {author} {\bibfnamefont {S.}~\bibnamefont {Bonelli}}, \bibinfo
  {author} {\bibfnamefont {F.}~\bibnamefont {Nicot}},\ and\ \bibinfo {author}
  {\bibfnamefont {F.}~\bibnamefont {Darve}},\ }\bibfield  {title} {\bibinfo
  {title} {Macroscopic softening in granular materials from a mesoscale
  perspective},\ }\href {https://doi.org/10.1016/j.ijsolstr.2020.02.022}
  {\bibfield  {journal} {\bibinfo  {journal} {International Journal of Solids
  and Structures}\ }\textbf {\bibinfo {volume} {193-194}},\ \bibinfo {pages}
  {222} (\bibinfo {year} {2020})}\BibitemShut {NoStop}%
\bibitem [{vid(2024)}]{video}%
  \BibitemOpen
  \href {https://youtu.be/L7UVJuSdv1g} {\bibinfo {title}
  {See Supplementary video}} \BibitemShut
  {NoStop}%
\bibitem [{\citenamefont {Cantor}\ \emph {et~al.}(2020)\citenamefont {Cantor},
  \citenamefont {Az{\'e}ma},\ and\ \citenamefont
  {Preechawuttipong}}]{cantorMicrostructuralAnalysisSheared2020}%
  \BibitemOpen
  \bibfield  {author} {\bibinfo {author} {\bibfnamefont {D.}~\bibnamefont
  {Cantor}}, \bibinfo {author} {\bibfnamefont {E.}~\bibnamefont {Az{\'e}ma}},\
  and\ \bibinfo {author} {\bibfnamefont {I.}~\bibnamefont {Preechawuttipong}},\
  }\bibfield  {title} {\bibinfo {title} {Microstructural analysis of sheared
  polydisperse polyhedral grains},\ }\href
  {https://doi.org/10.1103/PhysRevE.101.062901} {\bibfield  {journal} {\bibinfo
   {journal} {Physical Review E}\ }\textbf {\bibinfo {volume} {101}},\ \bibinfo
  {pages} {062901} (\bibinfo {year} {2020})}\BibitemShut {NoStop}%
\bibitem [{\citenamefont {{C{\'a}rdenas-Barrantes}}\ \emph
  {et~al.}(2022)\citenamefont {{C{\'a}rdenas-Barrantes}}, \citenamefont
  {Bar{\'e}s}, \citenamefont {Renouf},\ and\ \citenamefont
  {Az{\'e}ma}}]{cardenasExperimentalValidationMicromechanically2022}%
  \BibitemOpen
  \bibfield  {author} {\bibinfo {author} {\bibfnamefont {M.}~\bibnamefont
  {{C{\'a}rdenas-Barrantes}}}, \bibinfo {author} {\bibfnamefont
  {J.}~\bibnamefont {Bar{\'e}s}}, \bibinfo {author} {\bibfnamefont
  {M.}~\bibnamefont {Renouf}},\ and\ \bibinfo {author} {\bibfnamefont
  {{\'E}.}~\bibnamefont {Az{\'e}ma}},\ }\bibfield  {title} {\bibinfo {title}
  {Experimental validation of a micromechanically based compaction law for
  mixtures of soft and hard grains},\ }\href
  {https://doi.org/10.1103/PhysRevE.106.L022901} {\bibfield  {journal}
  {\bibinfo  {journal} {Physical Review E}\ }\textbf {\bibinfo {volume}
  {106}},\ \bibinfo {pages} {L022901} (\bibinfo {year} {2022})}\BibitemShut
  {NoStop}%
\end{thebibliography}
\end{document}